\documentstyle[aps,twocolumn,epsfig]{revtex} 
\begin{document}  
\draft  
\preprint{} 
\twocolumn[\hsize\textwidth\columnwidth\hsize\csname@twocolumnfalse%
\endcsname 
 
\title{ 
Fractal Stability Border in Plane Couette Flow 
} 
 
\author{Armin Schmiegel and Bruno Eckhardt} 
\address{Fachbereich Physik und Institut f\"ur Chemie und Biologie des  
Meeres,\\ der C.v.~Ossietzky Universit\"at, 
Postfach 25 03, D-26111 Oldenburg, Germany\\ 
and 
}  
\address{
Fachbereich Physik der Philipps Universt\"at 
Marburg, D-35032 Marburg, Germany\\ 
(permanent)} 
 
\date{\today} 
\maketitle 
\begin{abstract} 
We study the dynamics of localised perturbations in plane Couette flow 
with periodic lateral boundary conditions. For small Reynolds number  
and small amplitude of the initial state  the perturbation decays on a  
viscous time scale $t \propto Re$. For Reynolds number larger than  
about $200$, chaotic transients appear with life times longer than  
the viscous one. Depending on the type of the perturbation isolated initial 
conditions with infinite life time appear for Reynolds numbers larger than 
about $270$--$320$. In this third regime,  
the life time as a function of Reynolds number and amplitude is fractal.  
These results suggest that in the transition region the turbulent  
dynamics is characterised by a chaotic repeller rather than an attractor. 
\end{abstract} 
  

]

For a variety of simple flows linear stability theory predicts 
and experiments verify that the transition to turbulence proceeds 
by a sequence of instabilities at well defined critical values 
of the control parameter \cite{Chandrasekhar,Reid}. Prominent 
examples include a layer of fluid heated from below, flow between 
rotating cylinders, liquid jets, and stratified fluids. 
The transition to turbulence in pipe flow 
\cite{mass_flux_pipe_flow} or plane Couette flow  
\cite{subtransi_PCF,FinAmpPert} and to spiral turbulence in the flow between  
counter-rotating cylinders \cite{Coles} 
does not fit this pattern. If the Reynolds number is sufficiently 
large, turbulent dynamics can occur although the laminar profile 
is still stable. A formal linear  
stability analysis predicts either no instability (as in  
plane Couette flow \cite{Romanov}) or one for Reynolds  
numbers larger than the ones  
where experiments begin to show turbulent behavior  
(e.g. plane Poiseuille flow \cite{Transition_to_turbulence}). 
The nature of the transition is different from the cases with a  
linear instability. It depends on the size of the perturbation 
\cite{Acheson}, 
shows very strong intermittency \cite{Alstroem5.94,Alstroem8.94} and has no 
sharply defined stability border\cite{mass_flux_pipe_flow}. 
 
Such observations can be explained, if the bifurcations are 
sub critical, for then the new state extends to lower values of  
the Reynolds number and can be reached starting from finite amplitude 
perturbations \cite{Ginzburg_Landau_1,Ginzburg_Landau_2,Poiseuille_normal_form}.
Simple models, based on interacting wave vector 
triads or truncations of Galerkin systems, are compatible 
with this view.  
However, it has not been possible to follow these upper branches all  
the way down to where they 
(dis-)appear and to identify a critical  
Reynolds number. 
 
A different approach has recently focused on the eigenvectors of the  
linearised problem \cite{Rohr_I,Stab_ohne_Eigenwert,Chaos_trans,Farrell_Ioannou}.  
Very often the linearised hydrodynamic eigenvalue 
problem is not hermitian, so that the eigenvectors are not orthogonal. 
As a consequence, even for (negative) eigenvalues, small perturbations  
can be amplified a lot, so that the neglect of nonlinear interactions 
no longer can be justified. This explains how in the absence  
of linear instability the stability region around the 
laminar profile becomes small, perhaps irrelevantly small.  
One can imagine that because of this growth some of the turbulence  
seen in these systems is noise
induced \cite{Farrell_Ioannou}. 
 
To shed some more light on the dynamics in these cases, we have studied 
the evolution of perturbations in a numerical model for plane Couette  
flow. The numerical method was designed specifically to allow accurate 
long integration times. The life time of these perturbations was mapped  
out in an amplitude-Reynolds number plane. This allowed to identify three  
dynamical regimes and to study the transitions between them. The results 
indicate that there is no sharp transition to turbulence: the landscape 
of life times has fractal features, with some isolated life times longer  
than the finite integration time. Thus, whether an initial condition leads to  
an evolution identified as turbulent depends in a sensitive way on  
initial amplitude, Reynolds number and the experimental observation time.  
This is in qualitative agreement with experiments on plane Couette flow  
\cite{subtransi_PCF,FinAmpPert} and consistent with observations on pipe flow  
\cite{mass_flux_pipe_flow}. In addition, the numerical results 
suggest that the turbulent state belongs to a transient repeller rather  
than a turbulent attractor \cite{Crutchfield88a,Rohr_II}. 
 
The numerical model (see \cite{schmiegel3.95} for more details) is based on an  
expansion of the velocity field in Fourier and Legendre polynominals, 
\begin{equation} 
{\bf u}(x,y,z,t) =  
\sum_{{\bf k}, p} \tilde{\bf u}_{{\bf k}, p}(t) e^{i (k_x x + k_y y)} L_p(z)\,. 
\end{equation} 
The advantage of Legendre polynominals compared to Chebyshev polynominals 
is that they give rise to evolution equations which are energy 
conserving in the Eulerian, undriven limit. This is important for the 
long time evolution we want to study. 
For every velocity-component, we
used a set of $37$ Fourier modes 
on a hexagonal grid and $16$ Legendre-polynominals. 
The boundary conditions 
${\bf u}(x,z,\pm H/2)=0$ and continuity equation are included 
within the Lagrange formalism of the first kind\cite{schmiegel3.95}.
These restrictions finally leave 962 dynamically active
degrees of freedom.
 
The basic flow is normalised to 
have velocity $\pm U_0$ at $z=\pm H/2$. The Reynolds number is defined as 
$Re = U_0 H/2\nu$, with $\nu$ the kinematic viscosity. Lengths are 
measured in units of $H/2$ and time in units of $\tau=H/2U_0$.  
The Navier-Stokes equation for a perturbation 
${\bf u}$ of the initial profile ${\bf U}_0 = U_0 z {\bf e}_x$,  
\begin{eqnarray} 
\dot {\bf u} &=& - ({\bf u}\cdot\nabla) {\bf U}_0  
+ ({\bf U}_0\cdot\nabla) {\bf u} \nonumber\\ 
&\ & - ({\bf u}\cdot\nabla) {\bf u} 
- {1\over \rho} \nabla p + {1\over Re} \Delta {\bf u}\\ 
\nabla {\bf u} &=& 0 \, , 
\end{eqnarray} 
are replaced by nonlinear ordinary differential equations for the  
amplitudes $\tilde{\bf u}_{{\bf k},p}$. It was verified 
that the time evolution could be followed sufficiently accurately 
for scaled times up to $10000$. For $h=1\,cm$, the viscosity of 
water and a Reynolds number of $Re=400$ this corresponds to about 
$9\,min$. In experiments a localised initial disturbance spreads
and reaches the span wise walls within less than a minute
\cite{Experiments_on_transition,spotspreading}. The numerical
calculations thus extend to a time region where in experiments
the influence of the lateral walls are no longer negligible.
 
\begin{figure}
\begin{center}
\leavevmode
\epsfig{file=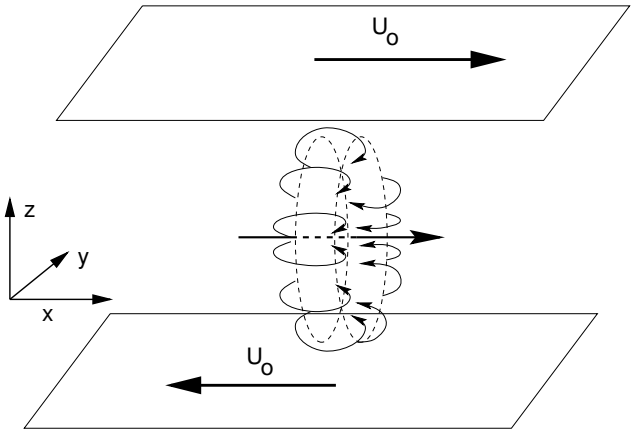,scale=.3}
\caption[]{Sketch of the polodial vortex-ring (\ref{initcond}) used
as initial condition.}
\label{init}
\end{center}
\end{figure}

The initial state for our simulations was taken to be a  
poloidal vortex ring in the $y$-, $z$-plane,  
\begin{equation} 
{\bf u} = A \, \mbox{curl}\,  \mbox{curl}\,  \delta(x,y) e^{-10 z^2}\, 
{\bf e}_x \, , 
\label{initcond}
\end{equation} 
with a variable amplitude $A$.
We approximate the $\delta$ function by setting all real Fourier modes 
equal to one.
As shown schematically in Fig.~\ref{init}, the axis of the ring points
in the ${\bf e}_x$-direction.  To remove excess energy for large modes 
we integrated this state for five time units at $\mbox{\rm Re}=400$ and used  
the resulting physically realistic initial condition in the subsequent 
studies. By this time the vortex ring has been rotated into the 
$x$-, $y$-plane and is similar to the  
one induced by the vertical water jets in the experiments 
of Berg\'e et al. \cite{subtransi_PCF,Experiments_on_transition,turbu_fluc_PCF}. 

\begin{figure}
\begin{center}
\leavevmode
\epsfig{file=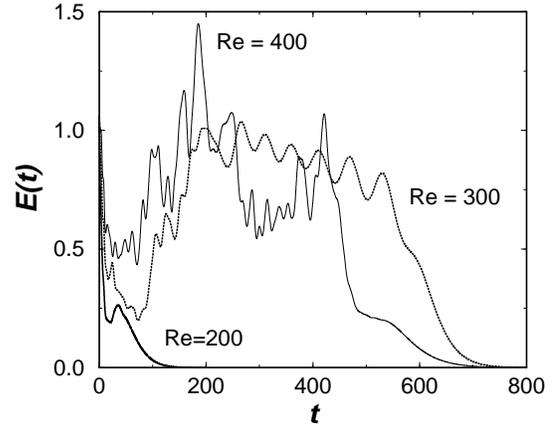,scale=.4}
\caption[]{Evolution of the energy in the pertubation {\em vs.} time
for three values of the Reynolds number.}
\label{tvsRe}
\end{center}
\end{figure}

Three different flow regimes
are distinguishable by their time series as shown in Fig.~\ref{tvsRe}.
For low Reynolds number, the energy of the perturbation decays   
smoothly, with a bump related to the non-normality of the 
linearised equations of motion. The time when this maximum 
is reached is in agreement with the estimate 
$t_{max} \propto 0.117\,Re$ in \cite{Energie_growth}.
This is the first regime, dominated by the linearised equations.

Around $\mbox{\rm Re}=300$ states show a qualitatively similar decay but only after an
intermediate period of irregular dynamics. A life time can be defined by
monitoring the $z$-component of the velocity until it decays below a certain
threshold. If that happens the perturbation lacks the driving and decays
as described by the linearised equations. In this second flow regime the
life times for perturbations are longer than estimated from the 
linearised equations. But one notes from Fig.~\ref{tvsRe}
that the life time of the disturbance is
not a monotonic function of $\mbox{\rm Re}$: The evolution at $\mbox{\rm Re}=400$ shows
more fluctuations and violent bursts 
but its life time is shorter than at $\mbox{\rm Re}=300$. 
In this third region, characterised by a non-monotonous variation 
of life times with amplitude and Reynolds number,
some initial conditions do not 
decay within the numerical integration time. The signal for the 
decaying and non-decaying states are dynamically similar in the 
fluctuating states, and there is no precursor that indicates 
the decay of the perturbation.

\begin{figure}
\begin{center}
\leavevmode
\epsfig{file=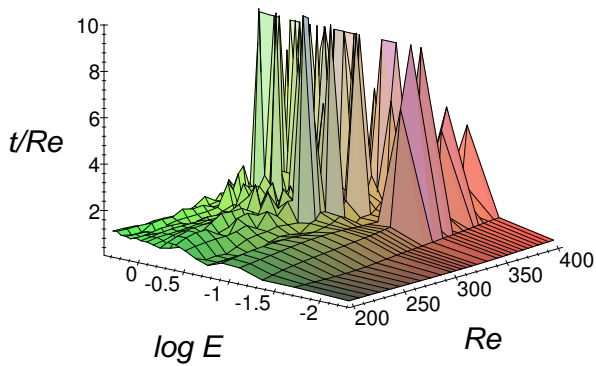,scale=.3,angle=-90}
\caption[]{
Lifetime {\em vs} Reynolds number $\mbox{\rm Re}$ and energy 
of the initial state. The life time is normalised by $\mbox{\rm Re}$,
so that in the linear viscous regime for small $\mbox{\rm Re}$ one expects
a constant. Integration was stopped after a scaled time of $3000$ units.
}
\label{life_time}
\end{center}
\end{figure}

The non-monotonous variations of life time with 
Reynolds number and amplitude occur for $\mbox{\rm Re}$ above about 350 and cover
an increasing range in amplitude. On a global scale (Fig.~\ref{life_time}), 
one notes a rather ragged landscape with a few peaks reaching  
up to the maximal life times followed in the numerical calculations. 
Because of the finite observation times in experiments states are classified
as decaying or turbulent according to the life times being above or below a 
certain threshold. Such a cut through the life time landscape shows
isolated turbulent disturbances in the decaying region and isolated decaying 
disturbances in the turbulent region. Such behavior has been
seen in experiments on constant mass flow through a pipe
\cite{mass_flux_pipe_flow}. 
The non-monotonicity observed there 
may thus be of dynamical origin and not due to experimental limitations. 

\begin{figure}
\begin{center}
\leavevmode
\epsfig{file=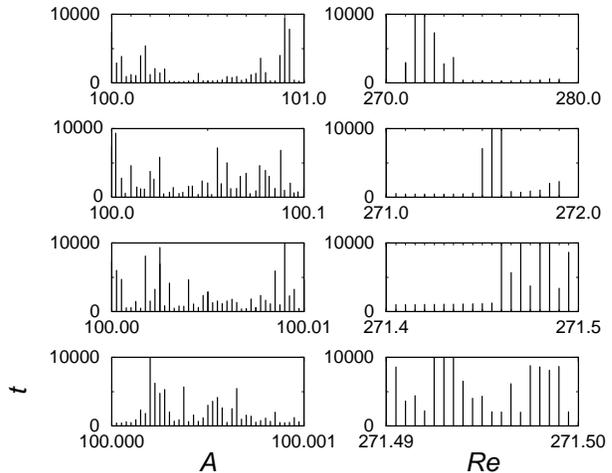,scale=0.3,angle=-90}
\caption[]{Successive magnifications of cuts through the life-time landscape 
of Fig.~\ref{life_time}, for $\mbox{\rm Re}=400$ and varying amplitude
(left column)
and for fixed amplitude $A=100$ and varying Reynolds number (right column).
}
\label{frac}
\end{center}
\end{figure}

The distribution of life times as a function of amplitude and 
Reynolds number has fractal features.  
Magnification of successive intervals in amplitude (Fig.~\ref{frac})  
for fixed Reynolds number or of intervals in $Re$ for 
fixed amplitude reveals a self similar pattern of life  
times, without any obvious simplifications on smaller scales. The  
magnifications are reminiscent of life time pictures obtained  
in chaotic scattering \cite{Eckhardt_Aref,Eckhardt,OttBuch}. 
This analogy suggests that the long life times arise for 
initial perturbations which are close to the stable manifold of 
saddles in phase space. Then arbitrarily large life times should
be possible and the `turbulent' state would be supported by 
a repeller rather than an attractor.

The magnifications in Fig.~\ref{frac} give no indication that the  
peaks with large life times have a finite width, in accord with the
repeller picture. The sensitivity of the life time to the initial
state can be quantified in terms of the uncertainty dimension
\cite{OttBuch}. 
For a given uncertainty in the amplitude $\epsilon$,
we calculate the probability $f(\epsilon)$ to find disturbance
with life times greater than a given limit $t_{\rm max} = 3000$.
For fractal sets, this probability decreases with decreasing $\epsilon$,
with a scaling exponent $\alpha$,  $f(\epsilon)\propto \epsilon^\alpha$. 
The  uncertainty dimension is then given by $d_s = 1-\alpha$. 
For $\mbox{\rm Re}=400$ we find a scaling exponent $\alpha \approx 0.29$ and 
an uncertainty dimension $d_s \approx 0.71$.

As the Reynolds number increases the number of long lived disturbances
increases and they cover an ever increasing region in phase space
and extend to smaller amplitudes. Phase space becomes more densely filled
with the stable and unstable manifolds of the saddles so that trajectories
trapped in this tangle need more and more time before they can escape
and relax to the laminar profile. This is in line with previous
observations in large systems \cite{Crutchfield88a} and  
pipe flow \cite{Brosa} that long lived transients can imitate 
a permanent turbulent state. Clearly a measure of the thickness of the 
repeller or an estimate of the life time of transients as a function
of Reynolds number would be highly desirable but is presently
beyond our numerical or analytical abilities.

Further support for these observations is derived from a low dimensional 
system of ordinary differential equations for a shear flow where we can 
follow the dynamics more efficiently and can also study the nature of the 
repeller in more detail \cite{Mersmann}. In particular, we find that the 
repeller is built around a rapidly increasing number of stationary 
states which are born in saddle-node bifurcations above a critical 
Reynolds number. All states are unstable and the dimension of the   
repeller is rather high-dimensional. In the present flow the 
time intervals on the  
repeller are too short and the dimensions presumably to high  
to allow for a determination of its fractal dimension. 
An indication of its dimension is given by the number of unstable 
eigenvalues of the local linearization. In our  
simulation of plane Couette flow at $\mbox{\rm Re}=400$ this number lies between $30$ and
$80$ positive eigenvalues. The maximum eigenvalue of the local linearization
was about $0.35$ in inverse time units.
 
Candidates for repeller states in plane Couette flow 
are the stationary states found by Nagata \cite{Bifurkation} 
and  Clever and Busse \cite{Clever_Busse_1,Clever_Busse_2}. 
These states appear for Reynolds numbers above about $125$. For larger
Reynolds numbers we have found further states, in accord with the 
observations on the model. But there remains a discrepancy between the
Reynolds number for the occurrence of stationary states (about 125) and
the one where the dynamics shows turbulence. However, the latter
depends strongly on the initial conditions.
For the ones presented above, the long lived states occur for $Re>300$, but 
if one takes poloidal rings with ${\bf e}_y$ instead of ${\bf e}_x$ 
they occur for $Re>270$ already. This strong dependence on the nature of the 
initial states indicates that the stable manifold of the repeller are 
curled up in the high-dimensional phase space. As the Reynolds number 
increases so does the region explored until it eventually overlaps 
with the initial state.  
 
To summarise, we have presented evidence for a fractal transition 
region to long lived turbulent states in plane Couette flow. 
These turbulent states are transient and belong to a repeller. 
The density of long life times increases with Reynolds number. 
This scenario might be relevant for other shear flows 
without linear instability and perhaps for some systems with 
a subcritical transition.  


\end{document}